\documentclass[letterpaper,prl,twocolumn,showpacs]{revtex4}

\usepackage{amsmath,amssymb}
\usepackage{epsfig}
\usepackage[geometry]{ifsym}

\newcommand {\ra}{\rightarrow}

\newcommand{\I}{\mathrm i}
\newcommand{\E}{\mathrm e}

\begin{document}
\title{On the nonlinear stage of Modulation Instability}
\date{\today}
\author{V.\,E.~Zakharov$^{1,2,3}$} \email{zakharov@math.arizona.edu}
\author{A.\,A.~Gelash$^{2}$} \email{agelash@gmail.com}
\affiliation{$^{1}$University of Arizona, Tucson, AZ, 857201, USA}
\affiliation{$^{2}$Novosibirsk State University, Novosibirsk, 630090, Russia}
\affiliation{$^{3}$Lebedev Physical Institute, Russian Academy of Sciences, Moscow, 119991, Russia}
\pacs{02.30.Ik, 05.45.Yv, 42.81.Dp, 47.35.Fg}
\begin{abstract}
We study the nonlinear stage of the modulation instability of a condensate in the framework of the focusing Nonlinear Schr\"{o}dinger Equation. We find a general $N$-solitonic solution of the focusing NLSE in the presence of a condensate by using the dressing method. We separate a special designated class of "regular solitonic solutions" that do not disturb phases of the condensate at infinity by coordinate. All regular solitonic solutions can be treated as localized perturbations of the condensate. We find an important class of "superregular solitonic solutions" which are small perturbations at certain a moment of time. They describe the nonlinear stage of the modulation instability of the condensate.
\end{abstract}
\maketitle {\it --Introduction --}
The focusing Nonlinear Schr\"{o}dinger Equation (NLSE) is a universal model for studying quasimonochromatic wave propagation in weakly nonlinear media. In particular it describes waves on deep water ~\cite{Zakharov1968}, waves in optical fibers ~\cite{Akhmediev-Ankiewicz2005}, Langmuir waves in plasma ~\cite{Sulem-Sulem1999}. Also this equation is a model of weakly interacting Bose gas with attraction between particles.

The NLSE has a simple solution, the monochromatic wave with frequency depended on amplitude - the condensate. The condensate is unstable with respect to modulation instability (The history of modulation instability is described in ~\cite{Zakharov-Ostrovsky2009}) What is a nonlinear stage of modulation instability? In spatial dimension $D=2,3$, the answer is known - modulation instability leads to formation of finite time singularities  - collapses ~\cite{Sulem-Sulem1999}.

In dimension $D=1$ collapses are forbidden. However in this case development of modulation instability leads to formation of "extreme" (rogue, freak) waves where energy density exceeds the mean level by order of magnitude. As a result, the study of long-time consequences of modulation instability is a problem of big practical importance, crucial for creation of a rogue wave theory in the ocean and a theory of extreme events in optical lines.

NLSE is a completely integrable system ~\cite{Zakharov-Shabat1972}, having many exact solutions. It is natural to hope that the nonlinear development of the modulation instability is described by such a solution.

In what follows we speak only about instability growing from a small spatially localized perturbation of condensate. Historically the first such solution was found by Peregrine in 1983 ~\cite{Peregrine1983}. This equation attracted a lot of attention ~\cite{Shrira-Geogjaev2010}. Its experimental confirmation was claimed ~\cite{Kibler2010,Chabchoub2011}. In 1985 a second order Peregrine solution was found ~\cite{Akhmediev-Elonskii-Kulagin1985}. Today "multi-Peregrine" are actively studied by different groups (see for instance ~\cite{Matveev2010}, ~\cite{Akhmediev-PRE2009}). All these solutions have a weak point  - they are small perturbations of the condensate only in the limit $t \ra -\infty$. These solutions are homoclinic - they describe freak waves appearing "from nowhere" and completely disappearing in the future. Meanwhile numerical modeling of modulation instability ~\cite{Agafontsev-Zakharov2012} as well as numerical solutions of exact Euler equation for potential flow of ideal deep fluid ~\cite{Zakharov-Dyachenko2010} demonstrate formation of propagating oscillating solutions ("breathers"). Thus the "homoclinic scenario" of modulation instability development can be disputed.

In this paper we present another class of exact solutions of NLSE which are small perturbations of condensate not at $t \ra \pm \infty$ but in the initial moment of time $t=0$. This is the special class of $2N$-solitonic solutions of NLSE in the presence of a condensate.

A solitonic solution of the NLSE in presence of condensate was first found by Kuznetsov in 1977 ~\cite{Kuznetsov1977} and rediscovered later in ~\cite{Ma1979,Kawata1978}. An important solitonic solution was found in 1985 by Akhmediev, Eleonskii and Kulagin ~\cite{Akhmediev-Elonskii-Kulagin1985} (the so called "Akhmediev breather"). This is a solution periodic in space and homoclinic in time. The Peregrine solution is the limiting case of both Kuznetsov's and Akhmediev's solutions. More general solitonic solutions were found in ~\cite{Its-Rybin-Sall1988, Tajiri-Watanabe1998, Akhmediev2009_PRA, Slunyaev2002, Zakharov-Gelash2011}. To construct exact solutions of NLSE describing the development of a localized small perturbation of the condensate one has use $2N$-solitonic solutions. There are several mathematical schemes for constructing such solutions (Hirota method, method of Darboux transformation etc.) We prefer the "dressing method"  elaborated in ~\cite{Zakharov-Mikhailov1978}. The details of our mathematical procedure can be found in ~\cite{Zakharov-Gelash2012}.

The main result of our parer is the following. One can construct a broad class of $2N$ - solitonic solutions depending on $7N$ parameters which form at $t=0$ a small perturbation of condensate. Evolution of this perturbation lead to the formation of complicated "integrable turbulence" ~\cite{Zakharov2009} where local concentration of energy easily exceeds in order of magnitude the energy density in the condensate.

We claim that all solitonic solutions in the presence of an unstable condensate, including the Peregrine and multi-Peregrine solutions, are automatically unstable. Our solutions are also unstable but we have constructed an infinite number of such solutions. They may be used as "bricks" for building up a consistent statistical theory of modulation instability development.

{\it --NLSE--}
We write the NLSE in the following form
\begin{equation}
\I\varphi_{t}-\frac{1}{2}\varphi_{xx}-(|\varphi|^{2}-|A|^{2})\varphi=0. \label{NLSE}
\end{equation}
The trivial condensate solution of equation (\ref{NLSE}) is $\varphi=\varphi_0=A$. One can consider $A$ to be real. This solution is unstable with respect to small perturbations. The growth rate of instability is $\Gamma(p)=p\sqrt{A^2-p^2/4}$, where $p$ is wave number of perturbation. In what follows we use the NLSE with non-vanishing boundary conditions $|\varphi|^2\rightarrow |A|^{2}$ at $x\rightarrow\pm\infty$. Equation (\ref{NLSE}) is the compatibility condition for the following overdetermined linear system for a matrix function $\mathbf{\Psi}$ ~\cite{Zakharov-Shabat1972}:
\begin{eqnarray}
\mathbf{\Psi}_x=\widehat{\mathbf{U}}\mathbf{\Psi}, \label{lax system 1}
\\
\I\mathbf{\Psi}_t=(\lambda\widehat{\mathbf{U}}+\widehat{\mathbf{W}})\mathbf{\Psi}. \label{lax system 2}
\end{eqnarray}
Here
\begin{eqnarray}
\widehat{\mathbf{U}}=\mathbf{I}\lambda+\mathbf{u},
\;\;\;\;\;\;\;\;
\widehat{\mathbf{W}}=\frac{1}{2}\begin{pmatrix}|\varphi|^{2}-A^{2} & \varphi_{x}\\ \varphi^*_{x} & -|\varphi|^{2}+A^{2}\end{pmatrix},
\label{U and W def. 1}
\nonumber\\
\mathbf{I}=\begin{pmatrix}1 & 0\\ 0 & -1\end{pmatrix},
\;\;\;\;\;\;\;\;
\mathbf{u}=\begin{pmatrix}0 & \varphi\\ -\varphi^* & 0\end{pmatrix}.
\label{U and W def. 2}
\end{eqnarray}
Here $\lambda$ is a spectral parameter. Suppose we know a certain particular solution $\varphi_0$ of equation (\ref{NLSE}) together with the fundamental matrix solution $\mathbf{\Psi}_0(x,t,\lambda)$ of system (\ref{lax system 1}), (\ref{lax system 2}). Then one can construct a new solution $\varphi$ of equation (\ref{NLSE}) using the following recipe. Choose $N$ complex numbers $\lambda_k$ (k=1,..N), $Re \lambda_k > 0$ and another set of arbitrary complex numbers $C_1,..C_n$. Denote $\mathbf{F}_k=\mathbf{\Psi}_0(x,t,-\lambda^*_k)$ and define $N$ vectors $\mathbf{q}_n$ by relation
\begin{eqnarray}
\mathbf{q}^*_n=\mathbf{F}_n
\left(
  \begin{array}{c}
    1 \\
    C_n \\
  \end{array}
\right).
\end{eqnarray}
Then a new solution is given by expression
\begin{equation}
\varphi=
\varphi_0+2 \widetilde{M}_{12}/M.
\label{N-solitoniic solution}
\end{equation}
Here $\widetilde{M}_{\alpha\beta}$ ($\alpha=1,2$) is the following determinant
\begin{equation}
\widetilde{M}_{\alpha\beta}=
\left|\begin{array}{cc}
        0 & \begin{array}{ccc}
              q_{1,\beta} & \cdots & q_{n,\beta}
            \end{array}
         \\
        \begin{array}{c}
          q^*_{1,\alpha} \\
          \vdots \\
          q^*_{n,\alpha}
        \end{array}
         &  \begin{array}{c}
              M^{T}_{nm}
            \end{array}
      \end{array}\right|.
\label{M1}
\end{equation}
Where $M_{nm}$ is a Hermitian matrix:
\begin{equation}
M_{nm}=\frac{(\mathbf{q}_{n}\cdot \mathbf{q}^*_{m})}{\lambda_{n}+\lambda^*_m},
\;\;\;\;\;\;\;\;\;\;\;\;
M=det(M_{nm}).
\end{equation}
Mention that transformation $\mathbf{q}_n \ra a_n \mathbf{q}_n$, where $a_n$ are arbitrary complex numbers, does not change the result of dressing. If $\varphi_0=A$
\begin{equation}
\mathbf{\Psi}_{0}(x,t,\lambda)=\begin{pmatrix}\E ^{\phi(x,t,\lambda)} & s(\lambda)\E ^{-\phi(x,t,\lambda)}\\ s(\lambda)\E ^{\phi(x,t,\lambda)} & \E ^{-\phi(x,t,\lambda)} \end{pmatrix}.
\label{condensat lax solution}
\end{equation}
Here
\begin{eqnarray}
\phi=kx+\Omega t ,\;\; k^{2}=\lambda^{2}-A^{2}
,\;\;
\Omega=-\I\lambda k ,\;\; s=-\frac{A}{\lambda+k}.
\nonumber
\end{eqnarray}
$\Psi_0$ has a cut from $-A$ to $A$. We perform the Jukowsky transform and map this plane onto the outer part of the circle of unit radius:
\begin{eqnarray}
\lambda=\frac{A}{2}(\xi+\frac{1}{\xi}),
\label{Uniformizing variable}
\end{eqnarray}
and use parametrization
\begin{eqnarray}
\xi_n=R_n \E^{\I\alpha_n}=\E^{z_n+\I\alpha_n},
\;\;\;\;\;\
C_n=e^{\I\theta_n+\mu_n}.
\end{eqnarray}
After redefinition of phase factor $\phi_n$
\begin{eqnarray}
q_{n1}=\exp(-\phi_n)+\exp(-z_n-\I\alpha_n+\phi_n),
\nonumber\\
q_{n2}=\exp(-z_n-\I\alpha_n-\phi_n)+\exp(\phi_n),
\nonumber\\
\phi_n=\ae_n x-\gamma_n t + \mu_n/2 + \I(k_n x-\omega_n t - \theta_n/2),
\nonumber\\
\ae_n=A\sinh z_n\cos\alpha_n,
\;\;\;\;\;\
k_n=A\cosh z_n\sin\alpha_n,
\nonumber\\
\gamma_n=-(A^2/2)\cosh 2z_n\sin 2\alpha_n,
\nonumber\\
\omega_n=(A^2/2)\sinh 2z_n\cos 2\alpha_n.
\label{qn vectors}
\end{eqnarray}
If $n=1$ we get a one-solitonic solution characterized by four parameters $R>1,\;\alpha,\;\theta,\;\mu$. The first two parameters define the location of the complex spectral parameter $\lambda$ which is actually a pole of the "dressing function" (see ~\cite{Zakharov-Gelash2012}). In the Kuznetsov case $\alpha=0,\;R>1$ and the pole is located on the real axis outside of the cut.

\begin{figure}[h]
\centering
\includegraphics[width=3in]{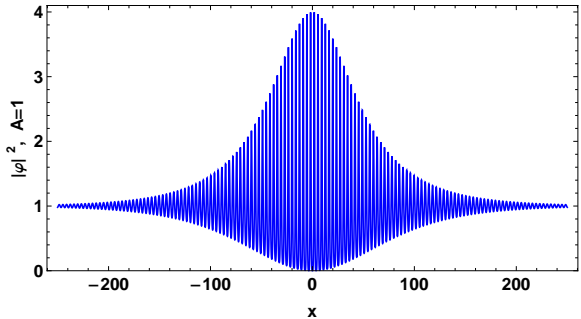}
\caption{\label{Near_Akhmediev}
The "quasi-Akhmediev" breather. $R=1.02,\alpha=\pi/3$.
}
\end{figure}

In the Akhmediev case the pole is located on the unit circle $R=1,\;\alpha \ne 0$. For the Peregrine solution $R=1,\;\alpha=0$. Now the pole is posed exactly in the branch point. In a general case $R>1,\;\alpha \ne 0$, the pole is located on the complex plan outside the unit circle. Parameters $\theta,\;\mu$ define the location and phase of the soliton. If $\theta=0,\;\mu=0$ the soliton satisfies the symmetry condition:
\begin{eqnarray}
\varphi(-x,-t)=\varphi^*(x,t).
\end{eqnarray}
In the general case the soliton is filled with carrying wave moving with phase velocity $V_{ph}$. The soliton's envelope moves with the group velocity $V_{gr}$
\begin{eqnarray}
V_{ph}=\omega/k,
\;\;\;\;\;\;\;\;\;\;\
V_{gr}=\gamma / \ae.
\end{eqnarray}
In the Kuznetsov and Peregrine case $V_{gr}=0$. In the Akhmediev case $V_{ph}=0,\;V_{gr}=\infty$. We are specifically interested in "quasi-Akhmediev" breather when the pole is close to the unit circle ($z<<1$). This is a quasi-periodic solution of large size $L \approx (zA\cos\alpha)^{-1}$. It moves with small phase and high group velocity:
\begin{eqnarray}
V_{ph} \approx \frac{Az\cos2\alpha}{\sin\alpha},
\;\;\;\;\;\;\;\;\;\;\
V_{gr} \approx -\frac{A\sin\alpha}{z}.
\end{eqnarray}
Note that the number of oscillations decreases with decreasing of $\alpha$. The quasi-Akhmediev breather is plotted on Fig.~\ref{Near_Akhmediev}.
The one-solitonic solution is defined by one vector $\mathbf{q}=(q_1,q_2)$. Its shape is given by the following formula
\begin{equation}
\varphi=
A
-
4A\cosh z\cos\alpha \;q^*_1 q_2/(|q_1|^2+|q_2|^2).
\label{Onesolitonic solution.form0}
\end{equation}
This solution has the following asymptotics
\begin{equation}
\varphi \ra A\exp(\pm 2\I\alpha)
\;\;\;\;\;\;\;\;\;\;\
x \ra \pm \infty.
\end{equation}
A general $N$-solitonic solution has asymptotics
\begin{eqnarray}
\varphi \ra A\exp\biggl(\pm2\I\sum^N_{k=1}\alpha_k \biggr)
\;\;\;\;\;\
x \ra \pm \infty.
\end{eqnarray}
If we are interested in $N$-solitonic solutions localized in a finite domain of space and not perturbing the remote condensate we must put
\begin{eqnarray}
\sum^N_{k=1}\alpha_k = 0,\; \pm \pi/2.
\end{eqnarray}
We call this solution a "regular solitonic solution" of the first ($\sum=0$) and second ($\sum=\pm \pi/2$) type. The difference between them is the direction of solitons movement. If we assume that the modulation instability develops from localized perturbation, only a regular solution can be used as a model for its nonlinear behavior. In what follows we are interesting only in regular solitonic solution of the first type.

Among one-solitonic solutions only the Kuznetsov and the Peregrine solutions are regular. In two-solitonic case we can construct a broad class of regular solutions.

\begin{figure}[h]
\centering
\includegraphics[width=3in]{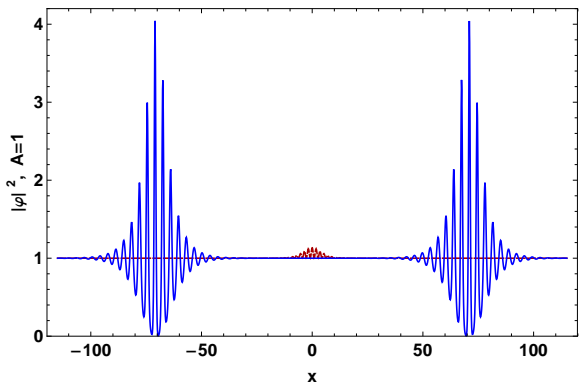}
\caption{\label{2S_SP}
The development of symmetric superregular two-solitonic solution. Red dashed line is small perturbation at the moment $t=0$. Blue solid line is the solution at the moment $t=15$. $\varepsilon=0.2,\;a=1,\;\alpha=\pi/3,\;\theta_1=\pi/2,\;\theta_2=\pi/2$.
}
\end{figure}

{\it --Two-solitonic solution--} The two solitonic solutions is defined by two vectors $\mathbf{q}_{1}=(q_{11},q_{12})$, $\mathbf{q}_{2}=(q_{21},q_{22})$. A general regular solitonic solution of the first type is depends on seven parameters $R_1,\;R_2,\;\alpha,\;\mu_1,\;\mu_2,\;\theta_1,\;\theta_2$. When $z_1=z_2=z$ poles located in complex conjugated points. In this case twosolitonic solution can be presented in the following form
\begin{eqnarray}
\varphi=A-A\sinh2z\sin2\alpha(N/ \Delta),
\nonumber\\
N=\sinh z\sin\alpha(|\mathbf{q}_1|^2 q^*_{21}q_{22}+|\mathbf{q}_2|^2 q^*_{11}q_{12})-
\nonumber\\
\I\cosh z\cos\alpha((\mathbf{q}^*_1 \cdot \mathbf{q}_2) q^*_{21}q_{12}-(\mathbf{q}_1 \cdot \mathbf{q}^*_2) q^*_{11}q_{22}),
\nonumber\\
\Delta=\cosh^2z\cos^2\alpha|q_{11}q_{22}-q_{12}q_{21}|^2+
\nonumber\\
\sinh^2z\sin^2\alpha|\mathbf{q}_1|^2|\mathbf{q}_2|^2.
\end{eqnarray}
Denote $\theta^{\pm} = \theta_1 \pm \theta_2$. Suppose that $\Delta \ne 0$ ($\theta^+ \ne 0$) and $z=0$. Then $\varphi = A$, thus the condensate is not perturbed. We call this phenomenon annihilation of Akhmediev breathers. If $z \ra 0$ the annihilation is "not complete". Solution is a small perturbation at the moment of solitons collision. Let us put $\mu_1=\mu_2=0$. In this case solitons collide at $(x=0,\;t=0)$. Let $R \simeq 1+\varepsilon $,\;$z \simeq \varepsilon$. Then solution is condensate solution plus small perturbation: $\varphi=A+\delta \varphi$. Perturbation $\delta \varphi$ is proportional to $\varepsilon$. When $\theta^+=\pi$ and $\exp(A^2 t \sin 2\alpha) < \varepsilon^{-1}$
\begin{eqnarray}
\delta\varphi \approx 4\I\varepsilon A N/\cosh(2A\varepsilon x \cos\alpha),
\nonumber\\
N=\cosh(A^2 t \sin 2\alpha + \I\alpha)\cos(2A x \sin\alpha-\theta^-/2).
\end{eqnarray}
This perturbations grow exponentially at the first time. An initially small localized perturbation of the condensate generates a pair of quasi-Akhmediev breathers propagating with very fast group velocity in opposite directions. Typical development of these small localized perturbation of the condensate is presented on Fig.~\ref{2S_SP}.

Annihilation of solitons takes place for much more general class of solitonic solutions. Let us consider a $2N$ solitonic solution consisting of $N$ pairs of Akhmediev breathers such that
\begin{eqnarray}
R_n=R_{n+N}=1,
\;\;\;\;\;\
\alpha_{}=-\alpha_{n+N},
\;\;\;\;\;\
n=1,...N.
\end{eqnarray}
Each pair is characterized by additional parameters $\mu_n,\;\mu_{n+N},\;\theta_n,\;\theta_{n+N}$. We assert that if all $\theta_n^+=\theta_n+\theta_{n+N} \ne 0$ this solution annihilates completely.
\begin{figure}[h]
\centering
\includegraphics[width=3in]{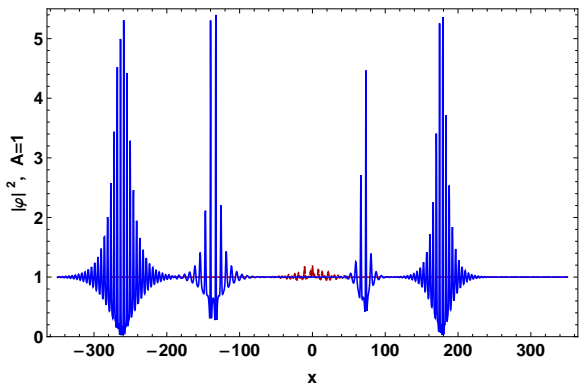}
\caption{\label{4S_SP}
The development of nonsymmetric superregular four-solitonic solution. Red dashed line is small perturbation at the moment $t=0$. Blue solid line is the solution at the moment $t=15$. $\varepsilon_1=0.05,\;a_{1}=1.5,\;\alpha_1=\pi/4,\;\theta_1=\pi/2,\;\theta_3=\pi/2$,
$\varepsilon_2=0.05,\;a_{2}=2,\;\alpha_2=\pi/7,\;\theta_2=\pi/2,\;\theta_4=\pi/2$.
}
\end{figure}
\begin{figure}[h]
\centering
\includegraphics[width=3in]{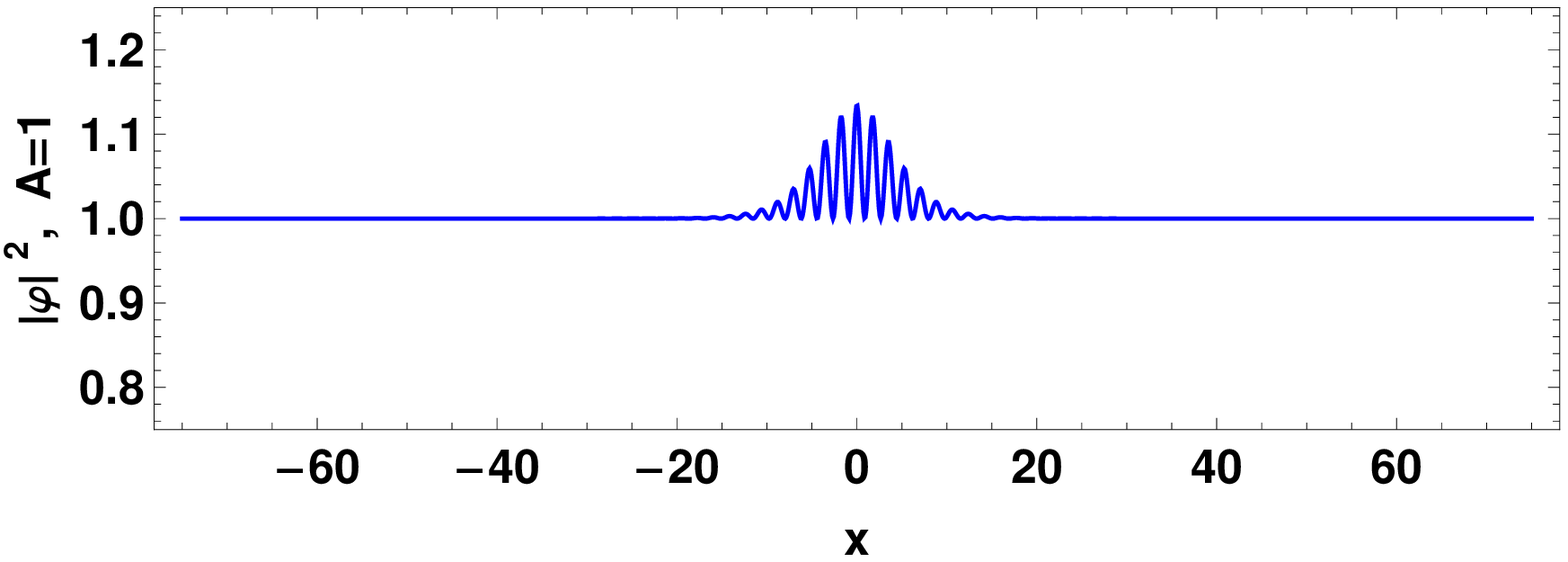}
\\
\includegraphics[width=3in]{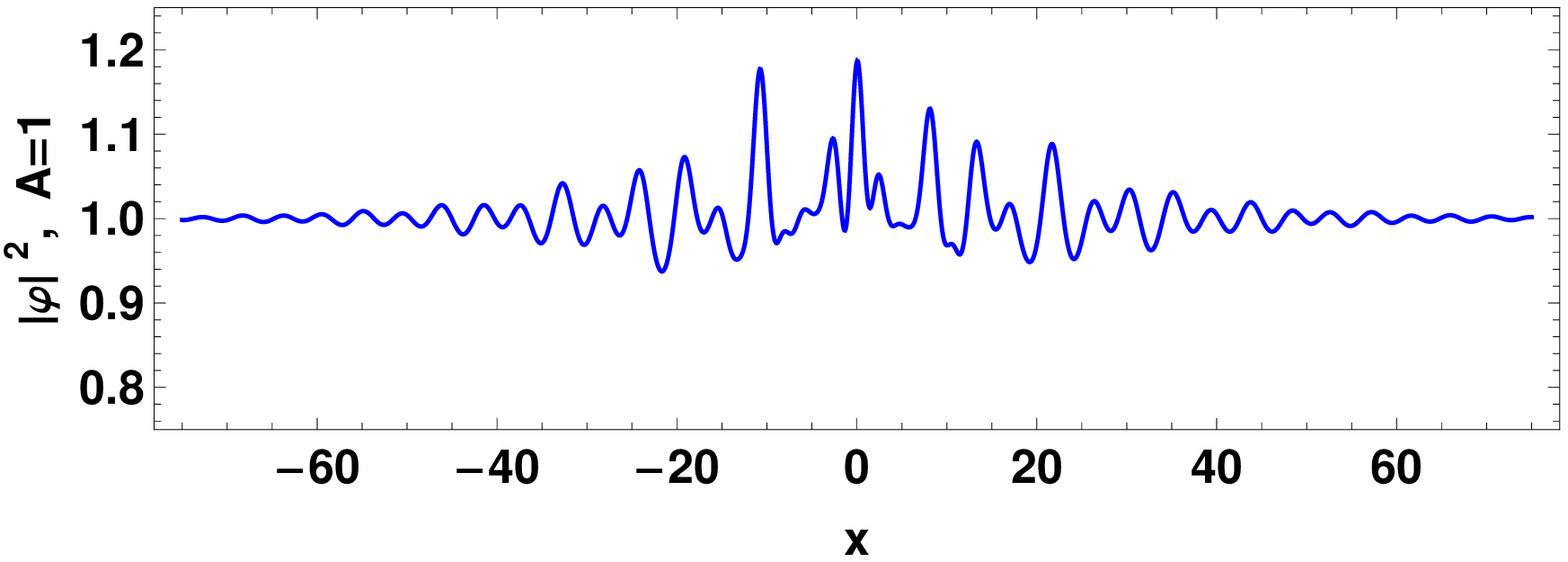}
\caption{\label{SP_zoom}
Enlarged small perturbations presented on Fig.~\ref{2S_SP} (top) and Fig.~\ref{4S_SP} (bottom).
}
\end{figure}
If the annihilation is "incomplete" such a solitonic solution presents a small perturbation of condensate. We call this solutions a "superregular solitonic solutions". Notice that an incomplete annihilation is not necessary symmetric. Even in the case $N=1$ one can put $R_1=1+\varepsilon$, $R_2=1+a\varepsilon$, $a \ne 1$, $\varepsilon \ra 0$. Such an initially small perturbation generates a pair of different near Akhmediev solitons propagating in opposite directions. A generic solution of the mentioned type is a nonlinear superposition of $N$ nonsymmetric pairs of quasi-Akhmediev breathers, and can be treated as a sort of "integrable turbulence" appearing as a result of nonlinear development of the modulation instability. The development of complicated nonsymmetric initial perturbation into four solitons is presented on Fig.~\ref{4S_SP}. What is interesting is that when $\varepsilon \ra 0$ this superposition is linear, so that at $t=0$ small perturbations generated by separate "superregular" pairs of soliton form an $N$ - dimensional linear space. This remarkable fact will be discussed in another article.

{\it --Conclusion --}
We have constructed a broad class of exact multisolitonic solutions of the NLSE describing localized in space and small at $t=0$ perturbations of the condensate. These solutions form an infinite-dimensional linear functional space. Most probably that any unstable localized  small perturbation of the condensate can be approximated by one of our solutions. In this case the "integrable turbulence" (see [25]) appearing as a result of the development of the modulation instability consists of quasi-Akhmediev breathers propagating in both directions. Self-consistent analytic theory of this turbulence will be a truly reliable theory of freak waves.
\begin{acknowledgments}
The authors express deep gratitude to Dr. E.A. Kuznetsov for helpful discussions.
This work was supported by: Russian Federation  Government  Grant (No. 11.G34.31.0035 with Ministry of Education and Science of RF, November 25, 2010), the RFBR (grant no. 12-01-00943), and by the Grant "Leading Scientific Schools of Russia" (No. 6170.2012.2).
\end{acknowledgments}

\end{document}